\documentclass[10pt]{iopart}

\usepackage{iopams}
\usepackage[T1]{fontenc}
\usepackage[utf8]{inputenc}
\usepackage{graphicx}
\usepackage{tabularx}
\usepackage{color}
\usepackage{textcomp}
\usepackage{enumerate}
\usepackage{todonotes}
\usepackage{hyperref}
\usepackage{comment}
\usepackage{url}
\usepackage{caption}
\usepackage{iopams}
\usepackage{cite}
\usepackage{url}
\usepackage{color,soul}
\usepackage{multirow}

\usepackage{xcolor}

\usepackage{todonotes}

\begin{document}

\title{A model comparison of 2D Cartesian and 2D axisymmetric models for positive streamer discharges in air}
\author{Zhen Wang$^{1,2}$, Anbang Sun$^{1}$, Jannis Teunissen$^{2}$}
\address{$^1$State Key Laboratory of Electrical Insulation and Power Equipment, School of Electrical Engineering,
  Xi'an Jiaotong University, Xi'an, 710049, China,
  $^2$Centrum Wiskunde \& Informatica, Amsterdam, The Netherlands\\
	}
\ead{jannis.teunissen@cwi.nl, anbang.sun@xjtu.edu.cn}

\begin{abstract}
  Simulating streamer discharges in 3D can computationally be very expensive, which is why 2D Cartesian simulations are sometimes used instead, especially when dealing with complex geometries.
  Although 2D Cartesian simulations can only be used to obtain qualitative results, it is nevertheless interesting to understand how they differ from their 3D or axisymmetric counterparts.
  We therefore compare 2D Cartesian and axisymmetric simulations of positive streamers in air, using a drift-diffusion-reaction fluid model with the local field approximation.
  With the same electrode length and width, inception voltages are found to be about a factor two higher in the 2D Cartesian case.
  When compared at the same applied voltage, the 2D Cartesian streamers are up to four times thinner and slower, their maximal electric field is about 30\% lower and their degree of ionization is about 65\% lower, with the largest differences occurring at the start of the discharge.
  When we compare at a similar ratio of applied voltage over inception voltage, velocities become rather similar, and so do the streamer radii at later propagation times.
  However, the maximal electric field in the 2D Cartesian case is then about 20-30\% lower, and the degree of ionization is about 40-50\% lower.
  Finally, we show that streamer branching cannot qualitatively be modeled in a 2D Cartesian geometry.


\end{abstract}

\maketitle

\ioptwocol

\section{Introduction}
Streamer discharges play an important role in the early stages of electric discharges~\cite{nijdam2020physics}, as they generate the first ionized paths that can later become leaders or sparks.
Due to their electric field enhancement, streamers can propagate into regions where the electric field is below the breakdown threshold of the insulating medium.
Streamer discharges often form a complex tree-like structure with many branched channels.
They are widely used in plasma and high voltage technology~\cite{fridman2005non,bruggeman2017foundations,Wang_2020}, and they appear in thunderstorms as streamer coronas ahead of lightning leaders or as sprite discharges high above thunderclouds~\cite{cummerSubmillisecondImagingSprite2006,mcharg2007observations,luqueEmergenceSpriteStreamers2009}.

Over the past decades, numerical simulations have been widely used to study streamer discharges.
Two types of models are commonly used, namely particle models~\cite{Rose_2011,teunissen20163d,Kolobov_2016,Levko_2017a,Stephens_2018a} and fluid models~\cite{Babaeva_2016,teunissen2017simulating,Plewa_2018,Marskar_2019a,Starikovskiy_2020,Ono_2020,wang2023quantitative}.
Simulations have been performed in different computational geometries.
3D Cartesian simulations are the most realistic, but also the most expensive.
A fine grid is required to accurately describe the thin charge layers and steep density gradients around streamer channels, and small time steps are required to describe the non-linear evolution of these channels.
Axisymmetric simulations are much cheaper to perform, as the solution only has to be evolved in two spatial coordinates $(r, z)$.
For single channels propagating in a straight line, identical results can be obtained as with a full 3D model, see e.g.~\cite{wang2022comparison}.
On the other hand, real streamer discharges are generally not axisymmetric, for example due to branching or other stochastic effects, or because of their interaction with dielectrics or electrodes.
In 2D Cartesian simulations, discharges evolve in $x, y$ coordinates while it is assumed there is no variation in the $z$-direction.
Such simulations therefore describe planar discharges with an infinite extent in the $z$-direction,
which can be a reasonable approximation for surface discharges that are approximately planar.


Because of the high cost of 3D simulations, 2D Cartesian simulations are often used to qualitatively study streamer discharges in complex geometries, which would in reality not be planar.
The goal of this paper is to be able to better interpret results of such 2D Cartesian simulations, by comparing them to axisymmetric simulations of positive streamers in air.
A planar discharge is expected to have weaker electric field enhancement, because its space charge layer is curved in only one instead of two dimensions.
We aim to understand how this difference affects streamer properties and the conditions for streamer inception.


Below, we briefly mention some of the past work on streamer discharges using 2D Cartesian simulations.
Such simulations have frequently been used to describe surface dielectric barrier discharges (SDBDs).
We remark that depending on the conditions, SDBDs can be approximately planar but they can also be highly filamentary, see e.g.~\cite{Sobota_2009,Stepanyan_2014,Ding_2022}.

Soloviev et al used a 2D Cartesian fluid model to study SDBDs in atmospheric air~\cite{soloviev2009surface, soloviev2014mechanism}.
Singh et al used a 2D Cartesian fluid model to simulate the propagation of streamer discharges towards and then along a solid surface, taking into account charge transport in the dielectric.
Meyer et al~\cite{meyer2019streamer} used a 2D planar fluid model to simulate positive surface streamers and the surface charge distribution on a grounded dielectric barrier.
The same model was used to study streamer propagation along a dielectric surface with a wave-like profile in~\cite{meyer2020streamer}.
In~\cite{li2020computationalpos,li2020computationalneg}, positive and negative streamers propagating over a dielectric surface were studied using a 2D Cartesian fluid model.
In~\cite{kruszelnicki2021propagation}, a two dimensional fluid model (nonPDPSIM) was used to simulate the propagation of discharges through interconnected pores in dielectric materials.
In~\cite{babaeva2018interaction}, the interaction of positive streamers in air with bubbles floating on liquid surfaces was computationally studied with a 2D Cartesian fluid model.


The outline of the paper is as follows.
In section \ref{sec:model-description}, the fluid models and the input data are described.
In section~\ref{subsec:comparison-same}, we compare 2D Cartesian and axisymmetric simulations of positive streamers in air at the same applied voltage.
Inception voltages are compared in section~\ref{subsec:inception}, and simulations are compared at different applied voltages in section~\ref{subsec:quantitative comparison}.
Finally, we discuss the effect of a 2D Cartesian geometry on streamer branching in section~\ref{subsec:branching}.

\section{Model description}
\label{sec:model-description}

Simulations are performed using Afivo-streamer, a code for drift-diffusion-reaction fluid simulations~\cite{teunissen2017simulating}.
The electron density evolves in time as
\begin{equation}
  \partial_t n_e =  \nabla \cdot (n_e\mu _e\mathbf{E} + D_e\nabla n_e) + S_i-S_a + S_\mathrm{ph}.
  \label{eq:fluid-model_ne}
\end{equation}
Here $\mu_e$ and $D_e$ are the electron mobility and diffusion coefficient, which are assumed to be functions of the local electric field. $S_\mathrm{ph}$ is a non-local photoionization source term discussed below, and $S_i$-$S_a$ are source terms due to ionization ($S_i$) and attachment ($S_a$) reactions, see section~\ref{subsec:input}.
Ions and neutral species are assumed to be immobile, and their densities $n_j$ (for $j = 1, 2, \dots$) evolve as
\begin{equation}
  \partial_t n_j =  S_j,
  \label{eq:fluid-model_ni}
\end{equation}
where the source terms $S_j$ are determined by the reaction list, see section~\ref{subsec:input}.
The electric field is computed as $\mathbf{E} = -\nabla \phi$, where the electric potential $\phi$ is obtained by solving Poisson’s equation using a parallel multigrid solver~\cite{teunissen2023geometric,teunissen2017simulating}.
The fluid equations are solved with a finite-volume method and explicit time integration, as described in~\cite{teunissen2017simulating}.

Photoionization is included according to Zhelenznyak's model~\cite{zhelezniak1982photoionization} using the Helmholtz approximation, using the three-term expansion given in \cite{bourdon2007efficient}.
The parameters used for photoionization are the same as those in~\cite{bagheri2019effect}.
In section~\ref{subsec:branching}, we additionally show some results with stochastic photoionization, which is implemented as a Monte-Carlo method with discrete photons, see~\cite{bagheri2019effect} for details.

The Afivo-streamer code includes adaptive mesh refinement (AMR), as described in~\cite{teunissen2017simulating,teunissen2018afivo}.
As a refinement criterion we use $\alpha(E) \Delta x < 1$, where $\Delta x$ is the grid spacing and $\alpha(E)$ is the field-dependent ionization coefficient.
The mesh is de-refined if $\alpha(E) \Delta x > 0.125$, but only if $\Delta x$ is smaller than 10 $\mu$m.


\subsection{Simulation conditions and computational domain}
\label{sec:initial-conditions}

\begin{figure}
  \centering
  \includegraphics[width=0.8\linewidth]{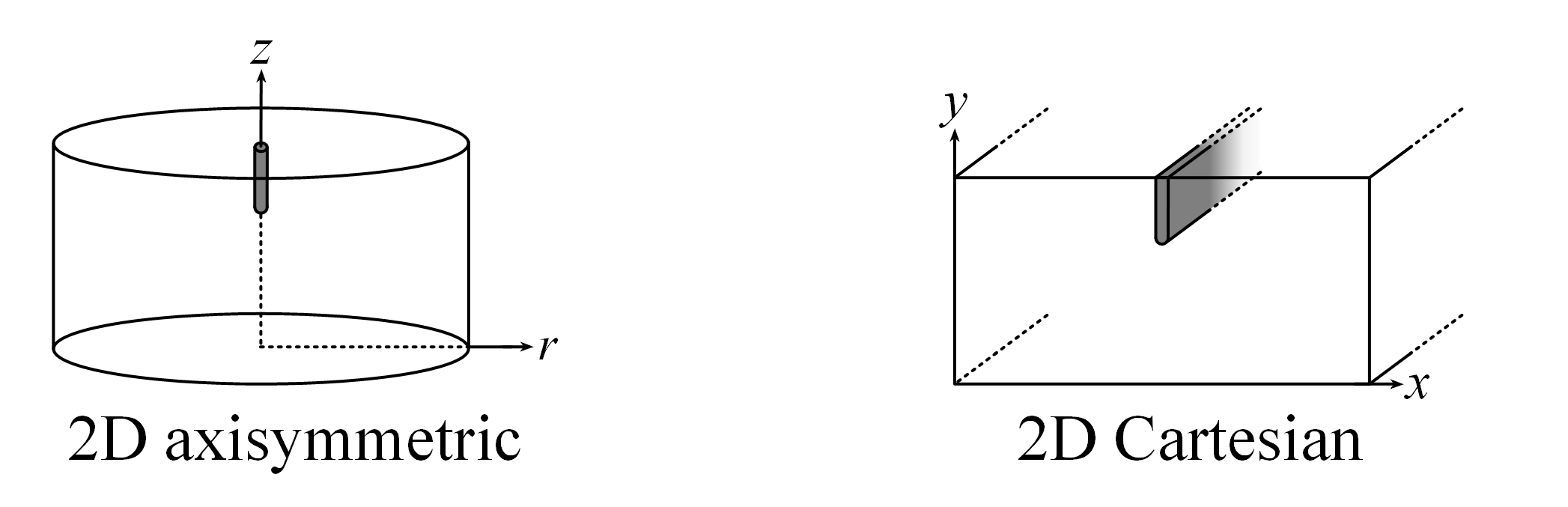}
  \includegraphics[width=\linewidth]{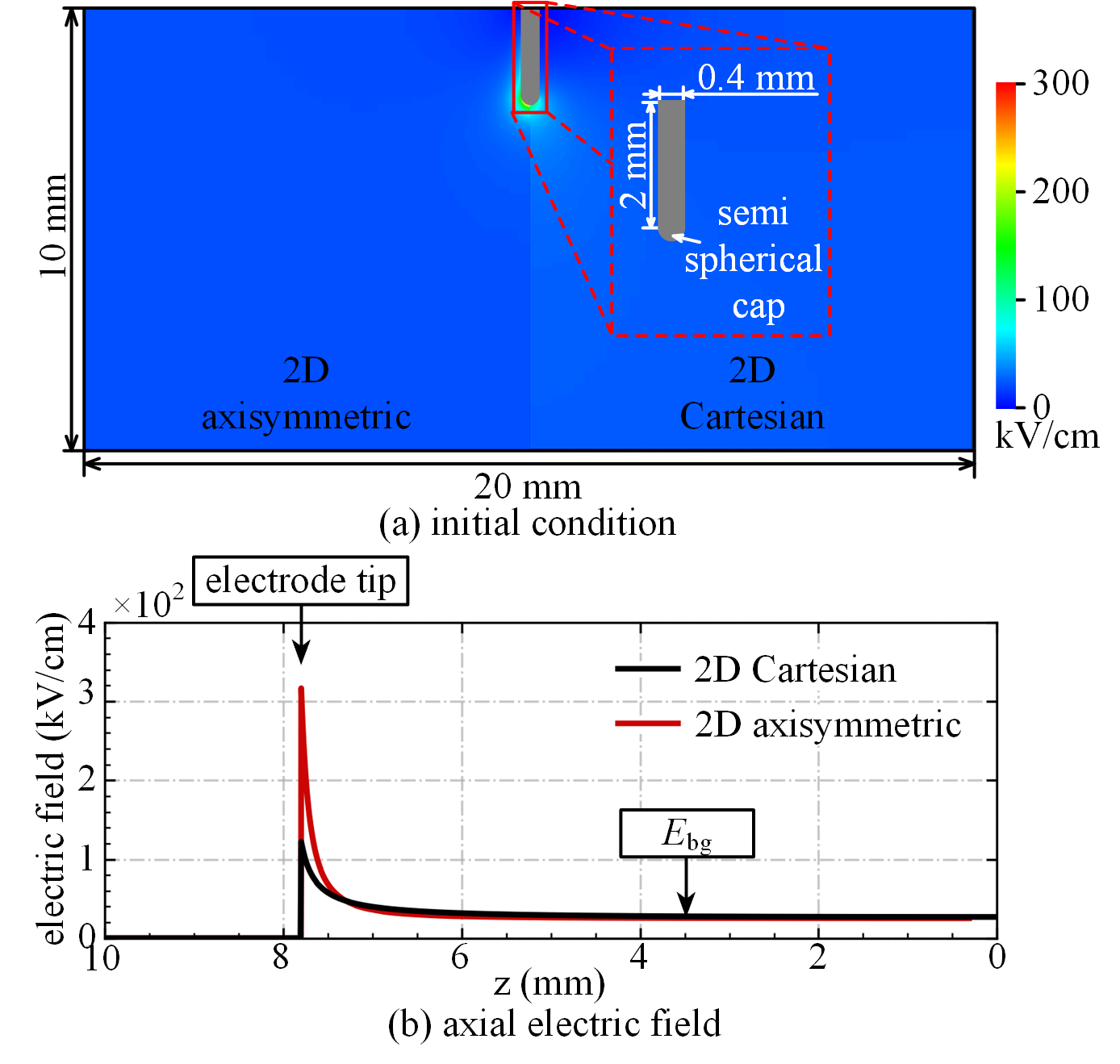}
  \caption{a) Illustration of the 2D Cartesian and 2D axisymmetric computational domains.
    b) The electrode and the electric field strength in the computational domains.
    c) Profile of the electric field strength along a central axis in the two models, for an applied voltage of 25\,kV/cm.
    $E_{\mathrm{bg}}$ is the average electric field between the plate electrodes.
  }
  \label{fig:comp-domain}
\end{figure}

Simulations are performed in artificial air (80\% N$_2$, 20\% O$_2$) at 1\,bar and 300\,K.
The computational domain used for the 2D Cartesian simulations measures $20~\mathrm{mm}\times10~\mathrm{mm}$, and a corresponding domain with a radius of $10~\mathrm{mm}$ and a height of $10~\mathrm{mm}$ is employed for the axisymmetric model, see figure~\ref{fig:comp-domain}.
We include a rod electrode with a semi-spherical cap to get electric field enhancement.
The electrode is $2~\mathrm{mm}$ long and has a radius of $0.2~\mathrm{mm}$.
Note that in the 2D Cartesian model, this rod actually becomes a blade-like electrode.


A Dirichlet boundary condition is used for the electric potential at upper and lower domain boundaries, and a homogeneous Neumann boundary condition is applied on the other boundaries.
Homogeneous Neumann boundary conditions are also used for species densities at all domain boundaries.

Figure~\ref{fig:comp-domain}(c) shows that the electric field enhancement at the electrode tip is about $2.6$ times higher in the 2D axisymmetric model, namely 310\,kV/cm compared to 120\,kV/cm.
This higher field decays more rapidly, so that farther away from the electrode the field is slightly higher in the 2D Cartesian model (the area under both curves is equal).
In both models, the electric field far away from the electrode is approximately equal to the average electric field $E_{\mathrm{bg}} = 25 \, \mathrm{kV/cm}$ between the plate electrodes.

As an initial condition, a homogeneous background ionization density of $1\times10^{10}~\mathrm{m}^{-3}$ electrons and positive ions is included, which provides initial free electrons so that a discharge can start.

\subsection{Input data}
\label{subsec:input}

\begin{table}
	\centering
	\caption{Reactions included in the model. Rate coefficients for $k_1$ to $k_5$ were computed using BOLSIG+~\cite{hagelaar2005solving,bolsig2019} from Phelps' cross sections~\cite{phelps1985anisotropic,Phelps-database}, and $k_6$ to $k_8$ were obtained from~\cite{Pancheshnyi_2005}.}
        \small
	\begin{tabular}{ll}
          \hline
          \hline
		Reaction & Rate coefficient \\ \hline
		$\textrm{e} + \textrm{N}_2 \stackrel{k_1}{\longrightarrow} \textrm{e} + \textrm{e} + \textrm{N}_2^+$ & $k_1(E/N)$ \\
		$\textrm{e} + \textrm{O}_2 \stackrel{k_2}{\longrightarrow} \textrm{e} + \textrm{e} + \textrm{O}_2^+$ & $k_2(E/N)$ \\
		$\textrm{e} + \textrm{O}_2 + \textrm{O}_2 \stackrel{k_3}{\longrightarrow} \textrm{O}_2^- + \textrm{O}_2$ 	& $k_3(E/N)$  \\
		$\textrm{e} + \textrm{O}_2 \stackrel{k_4}{\longrightarrow} \textrm{O}^- + \textrm{O}$ 			& $k_4(E/N)$  \\
		$\textrm{e} + \textrm{N}_2 \stackrel{k_5}{\longrightarrow} \textrm{e} + \textrm{N}_2(\textrm{C}^3 \Pi_u)$ 	& $k_5(E/N)$ \\
		$\textrm{N}_2(\textrm{C}^3 \Pi_u) + \textrm{N}_2 \stackrel{k_6}{\longrightarrow} \textrm{N}_2 + \textrm{N}_2$ & $k_6 = 0.13\times10^{-16}\,\textrm{m}^{3}\textrm{s}^{-1}$  \\
		$\textrm{N}_2(\textrm{C}^3 \Pi_u) + \textrm{O}_2 \stackrel{k_7}{\longrightarrow} \textrm{N}_2 + \textrm{O}_2$ & $k_7 = 3.0\times10^{-16}\,\textrm{m}^3\textrm{s}^{-1}$  \\
		$\textrm{N}_2(\textrm{C}^3 \Pi_u)\stackrel{k_8}{\longrightarrow} \textrm{N}_2(\textrm{B}^3 \Pi_g)$ 		& $k_8=1/(42\,\textrm{ns})$  \\
          \hline
          \hline
	\end{tabular}
	\label{tbl:reaction_table}
\end{table}

We use Phelps’ cross sections for N2 and O2~\cite{Lawton_1978,Pitchford_1982,phelps1985anisotropic,Phelps-database} and a relatively simple plasma chemistry, see table~\ref{tbl:reaction_table}.
Electron transport coefficients ($\mu_e and D_e$) and reaction rates are computed using BOLSIG+~\cite{bolsig2019,hagelaar2005solving}.
To get the optical radii of simulated streamers, we included the $\mathrm{N}_2(\mathrm{C}^3\Pi_u \to \mathrm{B}^3\Pi_g$) transition in the reaction list, since it is the main source of emitted light~\cite{pancheshnyi2000discharge} in $\mathrm{N}_2$-$\mathrm{O}_2$ mixtures close to atmospheric pressure.

\subsection{Computation of streamer radius}

\begin{figure}
  \centering
  \includegraphics[width=\linewidth]{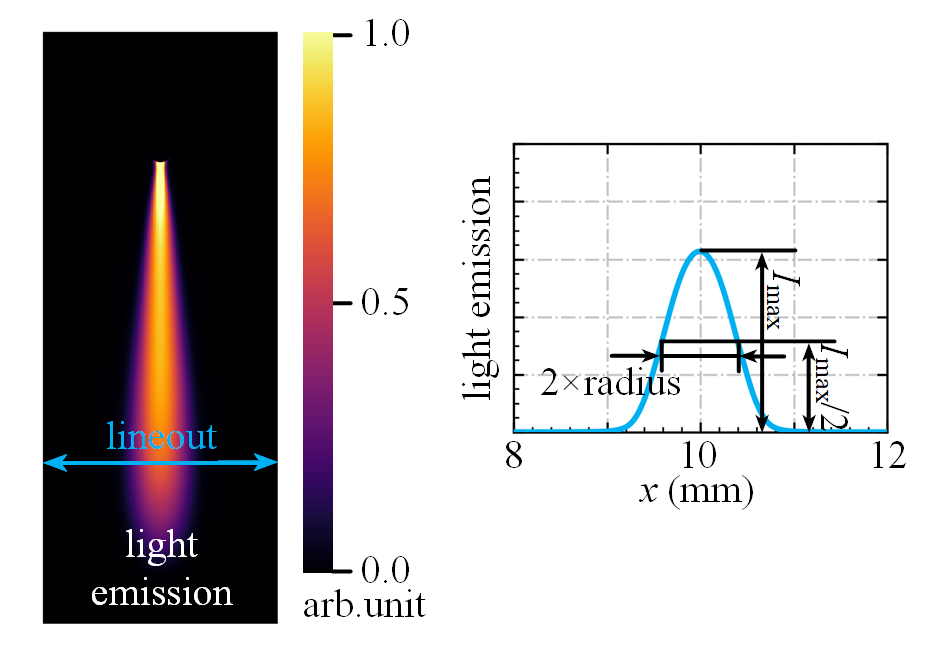}
  \caption{Illustration of how the streamer radius is determined from the light emission profile.
    At a given position, the full width at half maximum (FWHM) is determined by taking a lineout of the light emission, after which the radius is given by half of the FWHM.
  }
  \label{fig:radius-illustration}
\end{figure}

There are different definitions of the streamer radius.
In this paper we use the optical radius, defined as half of the FWHM (full width at half maximum) of the time-integrated light emission, see figure~\ref{fig:radius-illustration}.
For axisymmetric simulations, a forward Abel transform is first performed to compute the light emission profile as it would be observed experimentally.



\section{Results}
\label{sec:results}

\subsection{Comparison under the same applied voltage}
\label{subsec:comparison-same}

\begin{figure}
  \centering
  \includegraphics[width=\linewidth]{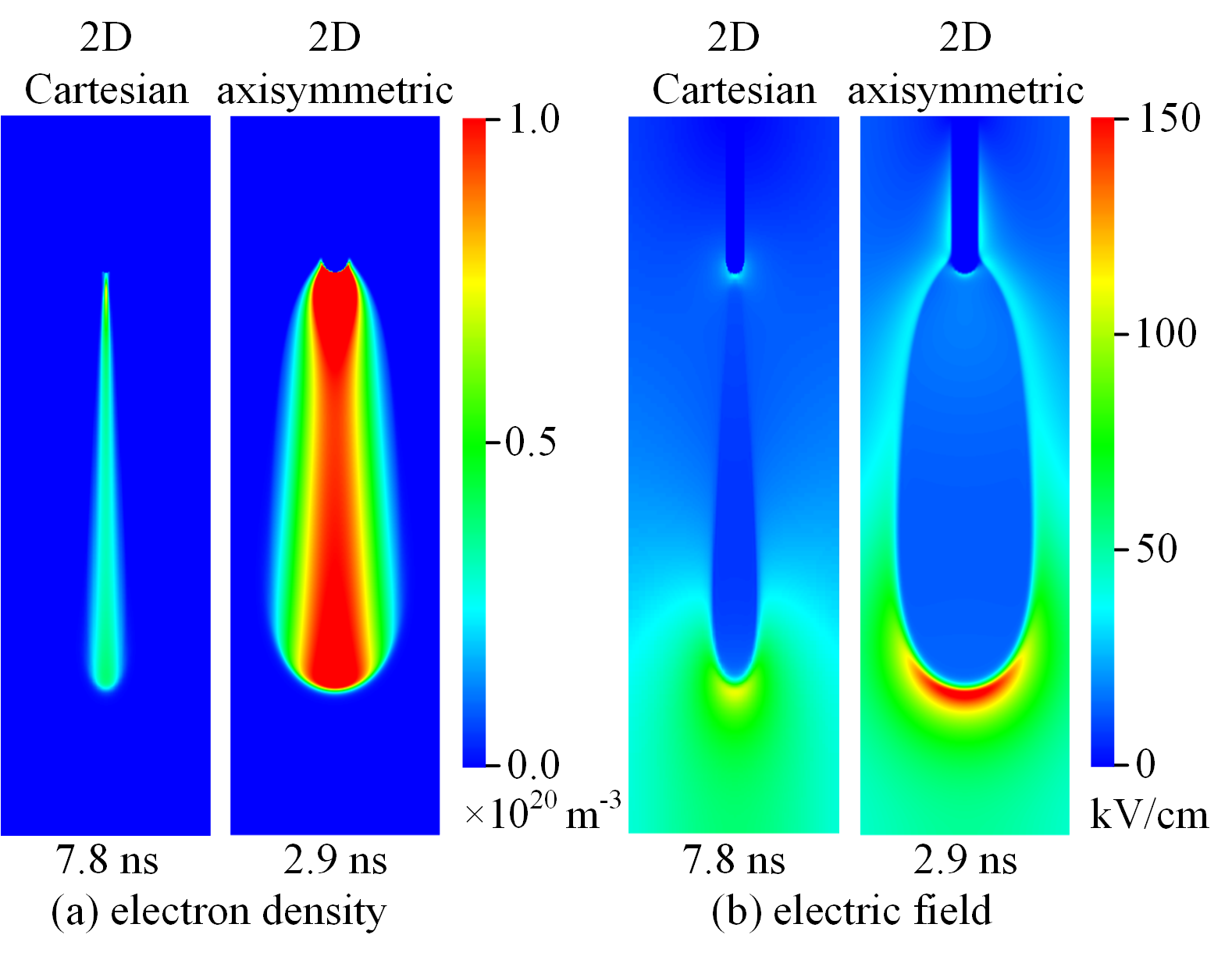}
  \caption{Comparison of electron density and electric field profiles for positive streamer discharges with an applied voltage of $U_0 = 25\,kV/cm$.
    Results are shown when the streamers approximately the same position in the the 2D Cartesian and axisymmetric simulations.}
  \label{fig:last-frame}
\end{figure}

\begin{figure}
  \centering
  \includegraphics[width=\linewidth]{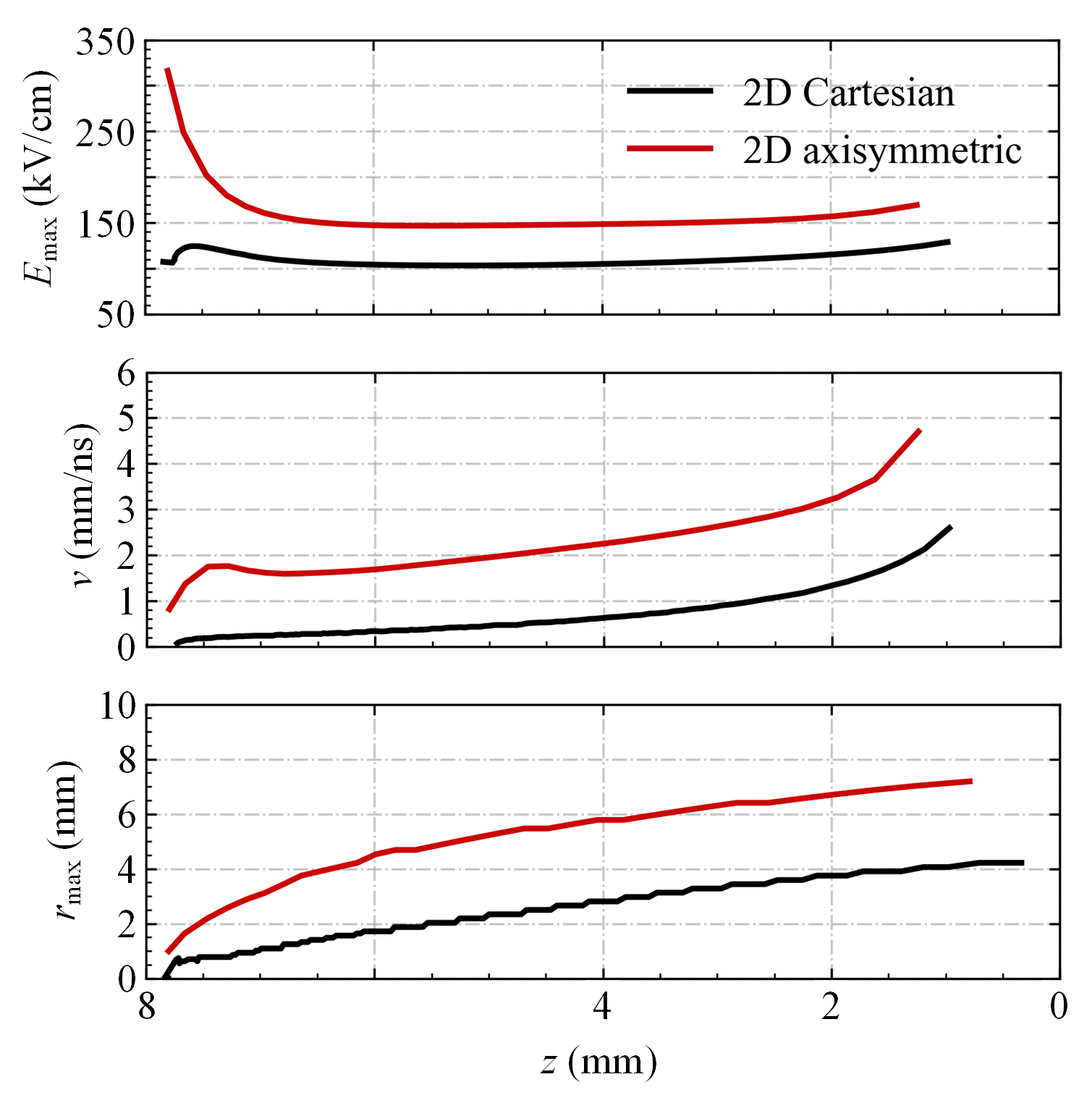}
  \caption{Comparison between 2D Cartesian and axisymmetric simulations for positive streamer discharges at an applied voltage of 25\,kV.
    The streamer head position is determined by the y/r-coordinate of the maximal electric field.
    Here a second order Savitzky–Golay filter of width five is used to compute a smoothed velocity from the streamer head position versus time data.}
  \label{fig:emax-velocity}
\end{figure}

In this section, 2D Cartesian and 2D axisymmetric simulations of positive streamer discharges are compared using the same applied voltage of $U_0 = 25~\mathrm{kV}$, which results in a background field of 25 kV/cm.
Figure~\ref{fig:last-frame} shows the electron density and the electric field from a 2D Cartesian simulation at $t=8.2~\mathrm{ns}$ and a 2D axisymmetric simulation at $t=3.2~\mathrm{ns}$.
Corresponding profiles of the maximal electric field, the streamer velocity and the streamer radius are shown in figure \ref{fig:emax-velocity}.

A much smaller streamer radius can be observed in the 2D Cartesian model, as well as a lower electric field at the streamer head and a lower channel conductivity.
Normally, one would expect a higher maximal field for a smaller streamer radius, so these differences are caused by the lower electric field enhancement in a 2D Cartesian geometry.
Both the radius and the velocity are initially already significantly higher in the axisymmetric model.
The maximal electric field relaxes to about $150 \, \textrm{kV/cm}$ in the axisymmetric model and to about $100 \, \textrm{kV/cm}$ in the 2D Cartesian model.

The two models clearly give rather different results when compared at the same applied voltage.
Due to the lower electric field enhancement in a 2D Cartesian geometry we have to use a relatively high voltage to get a discharge started.
In the axisymmetric simulations, this voltage is well above the inception voltage, leading to the formation of a wide and fast-propagating discharge channel.



\subsection{Comparison of inception voltage}
\label{subsec:inception}

In table~\ref{tab:incept-volt}, inception voltages are compared between the two models, defined as the lowest applied voltage that can initiate a streamer discharge.
The same computational domain as before is used, but we vary the electrode length and radius.
On average, inception voltages in the 2D Cartesian model are about two times higher than in the axisymmetric model.
Note that the inception voltage is somewhat more sensitive to the electrode geometry in the axisymmetric model due to the stronger field enhancement in this geometry.



\begin{table}
  \centering
  \caption{Streamer inception voltages $U_\mathrm{inc}$ for different electrode radii and electrode lengths.}
  \begin{tabular}{c|c|ccc}
    \hline
    \hline
    & electrode & \multicolumn{3}{c}{electrode radius}\\
    & length & $0.2$\,mm & $0.15$\,mm & $0.1$\,mm
    \\
    \hline
    \multirow{2}{*}{2D}
    & 2\,mm & 19.0\,kV    & 18.4\,kV    & 17.6\,kV \\
    & 3\,mm & 15.8\,kV      & 15.2\,kV    & 14.4\,kV \\
    \hline
    \multirow{2}{*}{axis.}
    & 2\,mm & 10.4\,kV    & 9.6\,kV    & 8.3\,kV \\
    & 3\,mm & 7.9\,kV     & 7.2\,kV    & 6.1\,kV \\
    \hline
    \hline
  \end{tabular}
  \label{tab:incept-volt}
\end{table}

\subsection{Comparison at different voltages}
\label{subsec:quantitative comparison}

We now compare 2D Cartesian and axisymmetric simulations at different applied voltages, considering two cases: a comparison at the same applied voltage $U_0$, and a comparison at a similar value of $U_0/U_\mathrm{inc}$, where $U_\mathrm{inc}$ is the inception voltage.
With the electrode geometry used in section~\ref{subsec:comparison-same}, $U_\mathrm{inc} = 19.0$\,kV for the 2D Cartesian model and $U_\mathrm{inc} = 10.4$\,kV for the axisymmetric model, see table~\ref{tab:incept-volt}.

\begin{figure*}
  \centering
  \includegraphics[width=\linewidth]{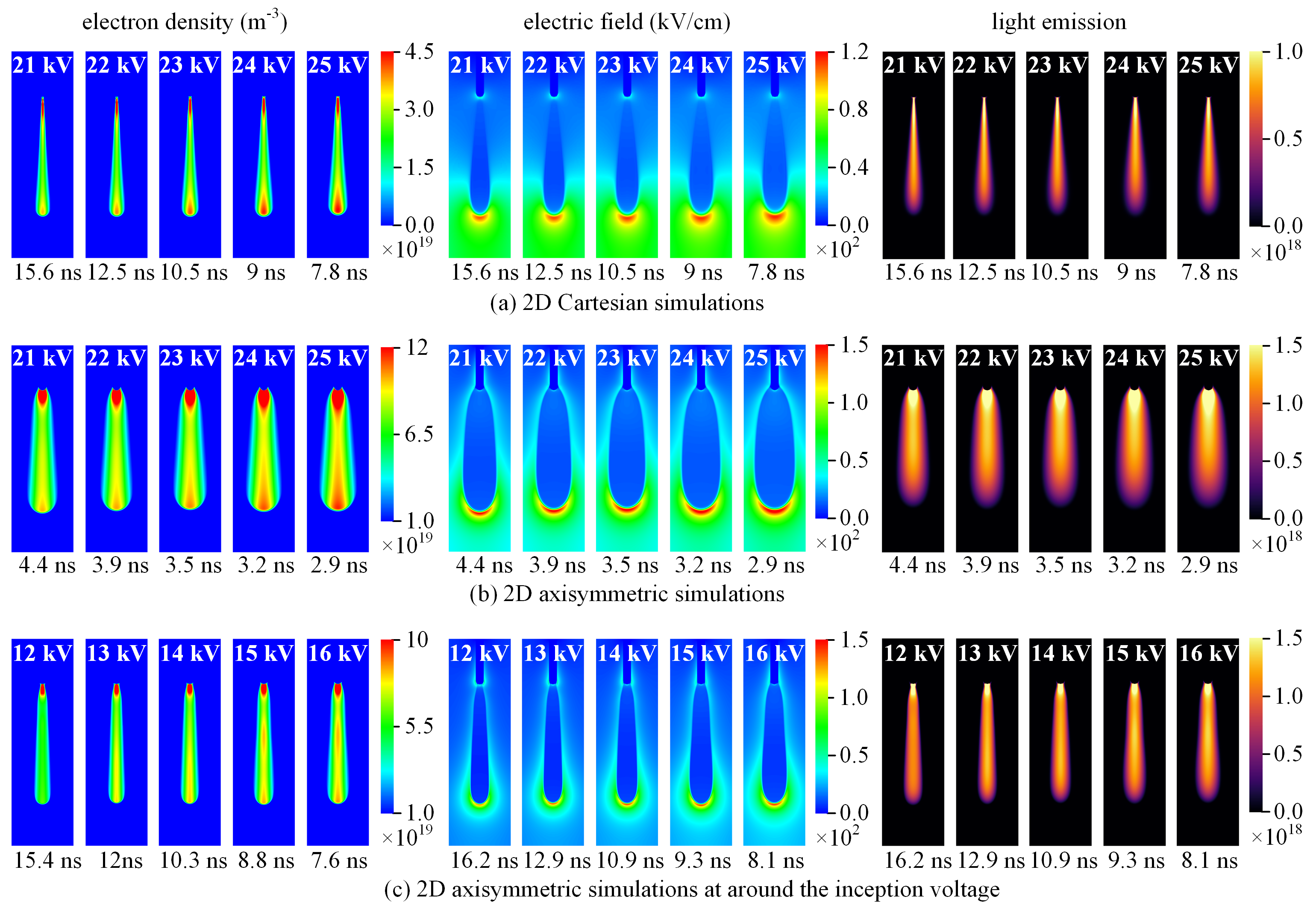}
  \caption{Comparison between 2D Cartesian and axisymmetric models under two groups of applied voltages, which are the group using the same applied voltages and the group using the voltages near the inception value.
    The electron density, background electric field, and light emission are presented from left to right.
    The presenting time is selected when the streamer tip reaches $z = 2~\mathrm{mm}$ The applied voltages are showed at the top of each plot and the time is at the bottom.
    Note that different color bars are used.}
  \label{fig:comparison}
\end{figure*}

\begin{figure*}
  \centering
  \includegraphics[width=\linewidth]{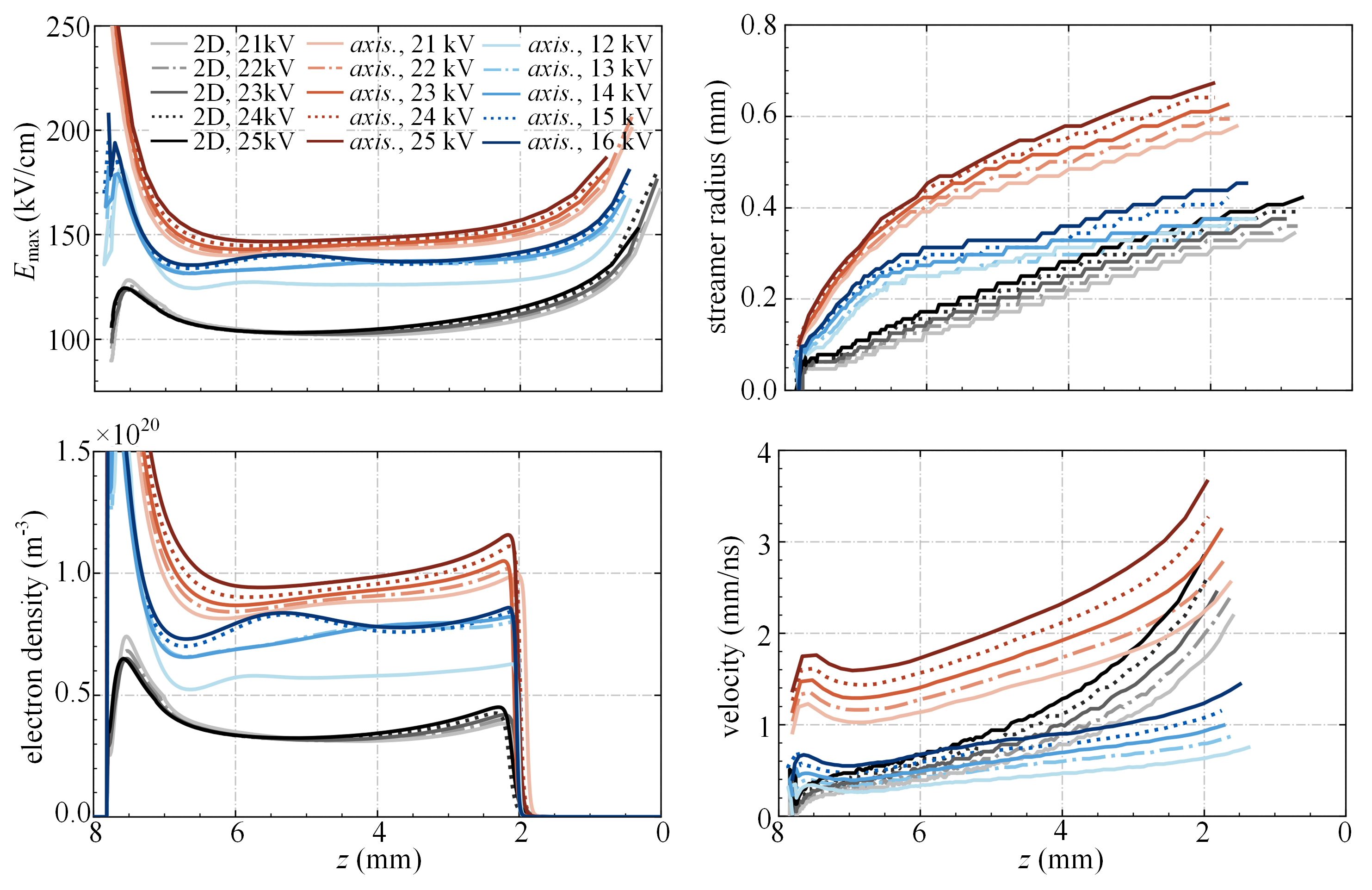}
  \caption{Streamer maximal electric field, radius and velocity versus the streamer head position, for all the cases shown in figure \ref{fig:comparison}.
    The streamer head position is defined as the vertical coordinate of $E_{\mathrm{max}}$.}
  \label{fig:properties-vs-pos}
\end{figure*}

Figure~\ref{fig:comparison} shows the electron density, electric field and light emission for different applied voltages at the moment the streamer heads reach $z = 2~\mathrm{mm}$.
In panels (a) and (b), the applied voltages are $U_0$ = 21, 22, 23, 24 and 25\,kV.
With these voltages, $U_0/U_\mathrm{inc}$ ranges from $1.11$ to $1.32$ for the 2D Cartesian model and from $2.0$ to $2.4$ for the axisymmetric model.
In panel (c) axisymmetric results are shown at lower voltages of 12, 13, 14, 15, 16\,kV, which correspond to $U_0/U_\mathrm{inc}$ ranging from $1.15$ to $1.54$.
Figure~\ref{fig:properties-vs-pos} shows the axial electron density profiles for all these cases, streamer radius, velocity and maximal electric field versus position.

When the 2D Cartesian simulations at 21\,kV--25\,kV are compared with the axisymmetric simulations at the same voltage, the same large differences as in section~\ref{subsec:comparison-same} are observed.
The axisymmetric streamer is two to four times wider, with the largest differences occurring near the electrode, and it is also two to four times faster.
Furthermore, the maximal electric field $E_\mathrm{max}$ is about 50\% higher in the axisymmetric case and the electron density in the channel is about two to three times higher, see figure~\ref{fig:properties-vs-pos}.

When the 2D Cartesian simulations are compared with the axisymmetric simulations at lower voltages, the streamers have similar velocities.
Their radii in the later stages of propagation are similar as well, although initially the 2D Cartesian streamers are much thinner.
However, several other differences persist.
For the axisymmetric streamers $E_\mathrm{max}$ is about 20--30\% higher and the electron density in the channel is about 50--100\% higher, see figure~\ref{fig:properties-vs-pos}.
Another important difference is in the decay of the electric field ahead of the streamer, which takes place over a longer distance in a 2D Cartesian geometry.
This effect can clearly be seen when comparing figures~\ref{fig:comparison}(a) and (c): the region with an electric field strength close to $E_\mathrm{max}$ is much smaller in the axisymmetric case.
This can explain why the 2D Cartesian streamers tend accelerate more rapidly when they approach the bottom electrode, as there is a larger region ahead of the discharge where the electric field exceeds the critical field.

\subsection{Relation between streamer properties}
\label{sec:properties-relation}

For streamers in air, several (mostly empirical) relations between streamer properties such as velocity, radius and maximal electric field have been established, see e.g.~\cite{Naidis_2009,nijdam2020physics}.
We now look at a couple of these relations to see whether they change in some particular way for planar (2D Cartesian) discharges.

\begin{figure}
  \centering
  \includegraphics[width=\linewidth]{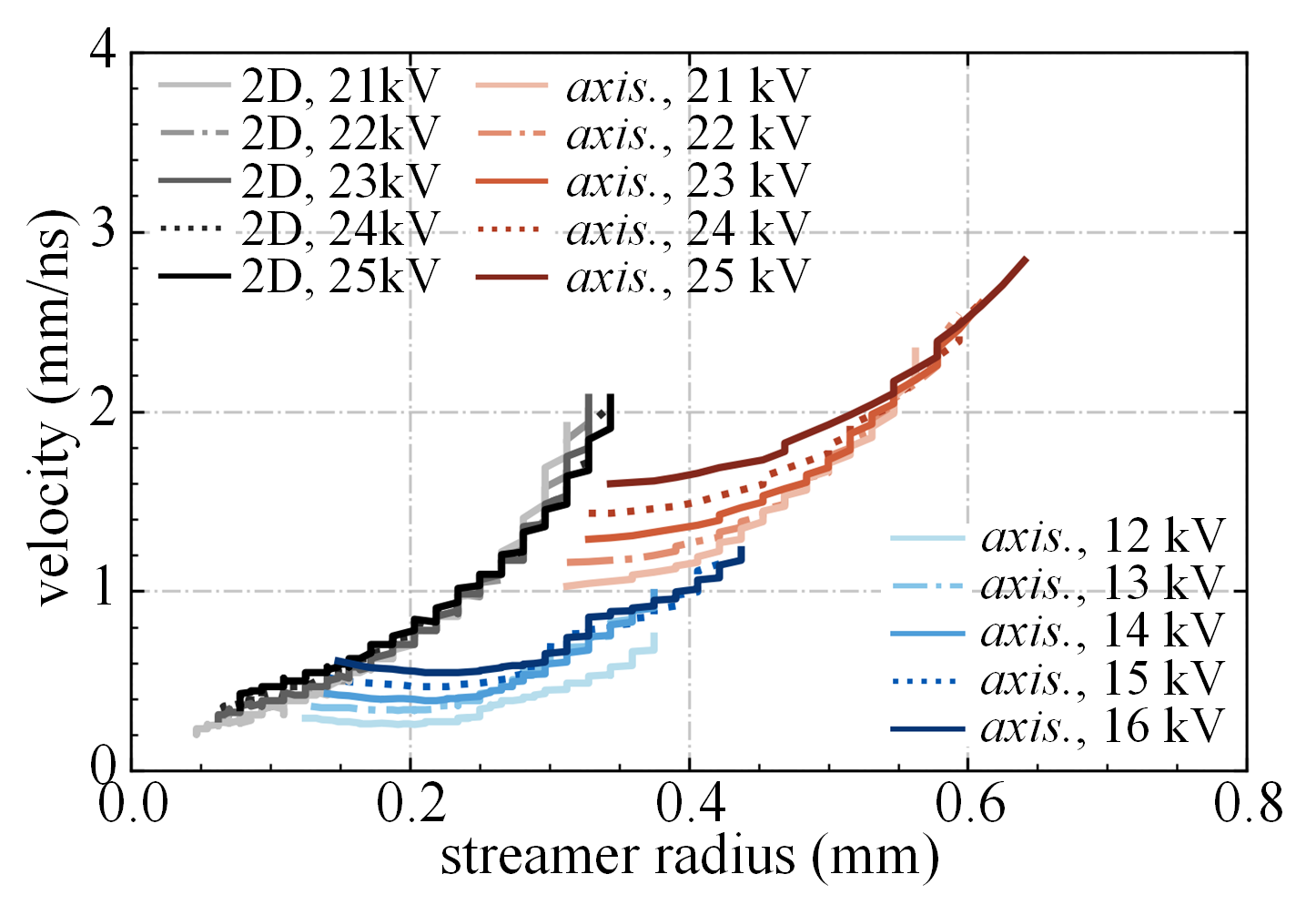}
  \caption{Streamer velocities $v$ versus the maximal streamer radius $r_\mathrm{max}$.
    The streamer radius is defined based on the light emission, using FWHM method.
    The same color scheme is used here as in figure~\ref{fig:properties-vs-pos}. The data are extracted when streamer head position is between 7-2~mm.}
  \label{fig:radius-velocity}
\end{figure}

Figure~\ref{fig:radius-velocity} shows the streamer radius versus velocity.
When the applied voltage is near the inception value for both models (21--25\,kV for the 2D Cartesian case and 12--16\,kV for the axisymmetric case), the velocities increase approximately linearly with the streamer radii.
When compared at the same streamer radius, the 2D Cartesian streamers typically have a higher velocity than the axisymmetric ones, but this is not surprising due to the difference in applied voltage.



Table~\ref{tab:alphaint} lists the electron density $n_e$ in the streamer channel (just behind the streamer head) and the maximal electric field strength $E_\mathrm{max}$ at the moment the streamers reach $z = 4 \, \mathrm{mm}$.
These values are compared with the ionization integral~\cite{ebert1996,babaeva1996,Naidis_1997a}
\begin{equation}
  \label{eq:alphaint}
  n_{\alpha}(E_\mathrm{max}) = \frac{\varepsilon_0}{e} \int_{0}^{E_\mathrm{max}} \alpha_\mathrm{eff}(E) \mathrm{d}E,
\end{equation}
and the ratio $n_e/n_\alpha$ is given.
This integral is accurate for one-dimensional ionization waves, in which the charge layers are not curved.
In axisymmetric or 3D simulations of positive streamers it has been observed that the ratio $n_e/n_\alpha$ is typically about two~\cite{Li_2022}, which was recently related to the role of the displacement current~\cite{Bouwman_2023}.
We find the same ratio $n_e/n_\alpha \approx 2$ for the 2D Cartesian case, even though this geometry lies `in between' a planar 1D ionization wave and a 3D streamer.
It is also interesting to note that $E_\mathrm{max}$ is not sensitive to the applied voltage for voltages between 21\,kV and 25\,kV.



\begin{table}
  \centering
  \begin{tabularx}{1.0\linewidth}{ccccc}
    \hline
    \hline
    Voltage & $E_\mathrm{max}$ & $n_e$ & $n_\alpha$ & $n_e/n_\alpha$\\
    (kV) & (kV/cm) & ($10^{19} \, \mathrm{m}^{-3}$) & ($10^{19} \, \mathrm{m}^{-3}$) & -\\
    \hline
    \multicolumn{5}{c}{2D Cartesian}\\
    21 & 102.1 & 3.16 & 1.46 & 2.16 \\
    22 & 102.6 & 3.18 & 1.49 & 2.14 \\
    23 & 103.3 & 3.22 & 1.53 & 2.11 \\
    24 & 104.3 & 3.29 & 1.58 & 2.08 \\
    25 & 105.3 & 3.38 & 1.64 & 2.06 \\
    \hline
    \multicolumn{5}{c}{Axisymmetric}\\
    21 & 143.3 & 8.95 & 4.70 & 1.90 \\
    22 & 144.5 & 9.12 & 4.83 & 1.89 \\
    23 & 145.7 & 9.34 & 4.96 & 1.88 \\
    24 & 147.1 & 9.57 & 5.11 & 1.87 \\
    25 & 148.7 & 9.85 & 5.29 & 1.86 \\
    \hline
    \multicolumn{5}{c}{Axisymmetric}\\
    12 & 126.1 & 5.99 & 3.09 & 1.94 \\
    13 & 136.8 & 7.97 & 4.05 & 1.97 \\
    14 & 137.1 & 7.89 & 4.08 & 1.93 \\
    15 & 136.2 & 7.93 & 3.99 & 1.99 \\
    16 & 137.3 & 8.06 & 4.09 & 1.97 \\
    \hline
    \hline
  \end{tabularx}
  \caption{Degree of ionization $n_e$, maximal electric field $E_\mathrm{max}$ and value of ionization integral $n_\alpha$, see equation~(\ref{eq:alphaint}).
    The values were obtained for the cases shown in figures~\ref{fig:comparison}--~\ref{fig:properties-vs-pos} when the streamers reached a vertical position $z = 4 \, \mathrm{mm}$.}
  \label{tab:alphaint}
\end{table}

\subsection{Branching}
\label{subsec:branching}

Branching determines the morphology of a streamer discharge tree and it influences streamer properties, for example because wider channels are more likely to branch than thinner ones.
In previous work~\cite{wang2023quantitative}, we have studied positive streamer branching in artificial air, using a 3D drift-diffusion-reaction fluid model coupled with stochastic photoionization, and we found good agreement with experimental observations.
In earlier work, streamer branching has also been computationally studied in a 2D coaxial geometry~\cite{Xiong_2014}.
An interesting question is therefore to what extent branching can qualitatively described in a 2D Cartesian geometry.

We expect branching to be significantly weaker in a 2D Cartesian geometry.
A first difference is that charge layers are only curved in one dimension.
This curvature drives the branching process through a Laplacian instability, since a protrusion can locally increase the electric field enhancement, see e.g.~\cite{Ebert_2011,nijdam2020physics}.
In a 2D Cartesian geometry this instability will be significantly weaker.
A second difference is that the electric field ahead of a 2D Cartesian streamer has a lower maximum but decays over a longer distance.
This leads to less steep electron density gradients, and therefore a reduced probability of branching.
A third difference is that higher applied voltages are required to initiate discharges in 2D Cartesian simulations.
Streamer can usually grow wider, and thus branch later, with a higher applied voltage.

\begin{figure}
        \centering
        \includegraphics[width=\linewidth]{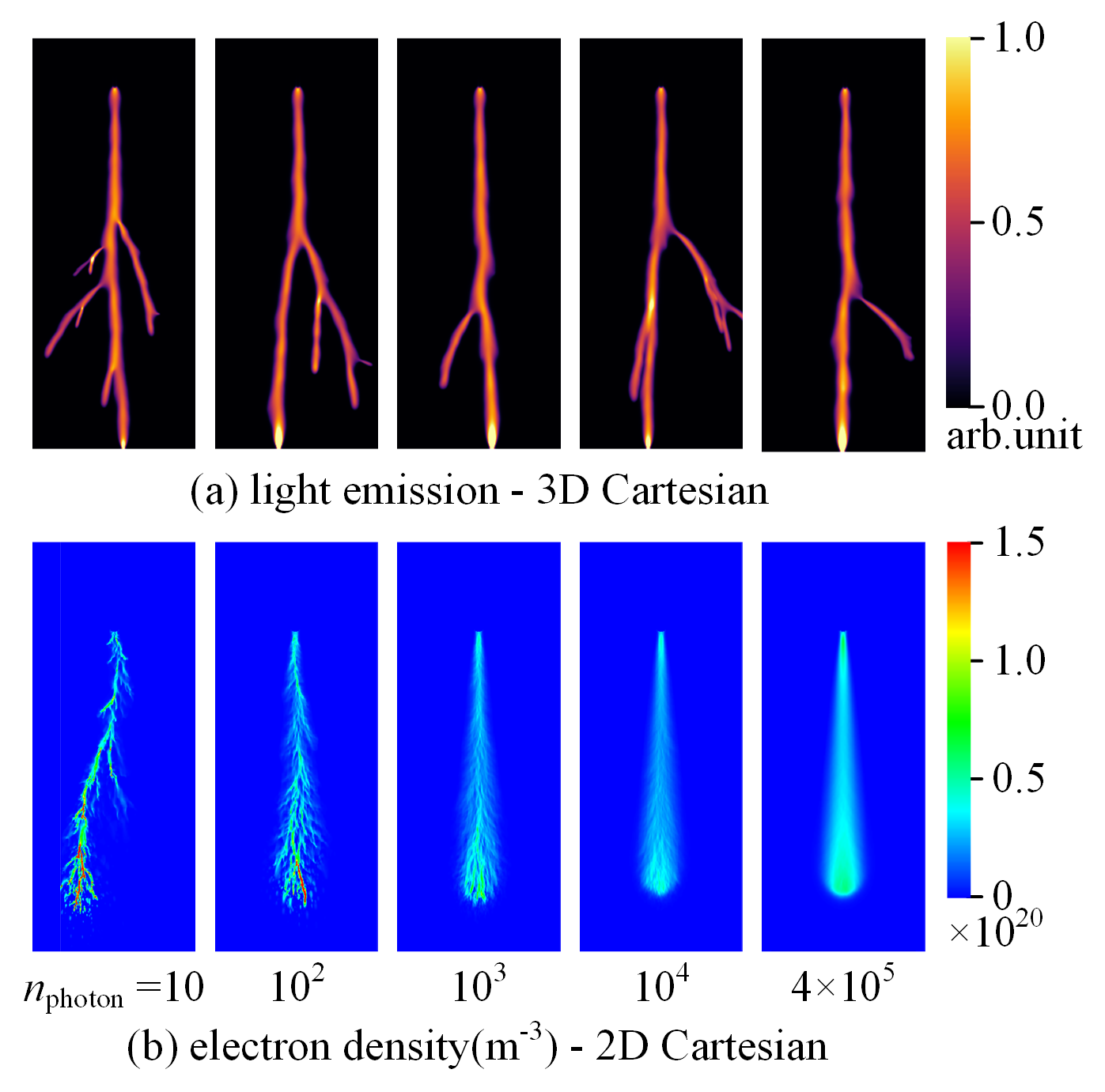}
         \caption{Examples of positive branching streamer in 2D Cartesian model and 3D Cartesian model. The stochastic photoionization is applied in both model to get streamer branching. 5 runs are showed for 3D Cartesian model under the same initial condition, see details in~\cite{wang2023quantitative}. Different desired weight of super-photons $n_\mathrm{photon}$ are discussed in 2D Cartesian model, and one run is showed for each value of $n_\mathrm{photon}$.}
         \label{fig:branching-example}
  \end{figure}

To test these ideas, we have performed 2D Cartesian simulations with a Monte Carlo photoionization model.
By limiting the maximum number of photons $n_\mathrm{photon}$ that are used to compute the photoionization source term (which is updated every time step), we can artificially increase the amount of noise and induce branching in this model.
Figure~\ref{fig:branching-example} shows examples of 3D Cartesian simulations from our previous work and examples of branching streamers in 2D Cartesian simulations for different values of $n_\mathrm{photon}$.
The branching streamers in the 2D Cartesian simulations have a feather-like shape with many very thin branches, which is completely different from the morphology observed in the 3D simulations and in real discharges in air.
This type of branching can only be induced for relatively small values of $n_\mathrm{photon}$, showing that branching is suppressed rather strongly.
We therefore conclude that 2D Cartesian models cannot qualitatively reproduce streamer branching.

\section{Conclusions}

In this paper, we have compared 2D Cartesian and 2D axisymmetric simulations of positive streamers.
The simulations were performed in air at 1\,bar and 300\,K, using a drift-diffusion-reaction fluid model.
An electrode of the same length and width was used in both geometries, corresponding to a needle in the axisymmetric case and a blade in the 2D Cartesian case.
The applied voltage was varied to obtain background fields ranging from 12\,kV/cm to 25\,kV/cm.

When compared at the same applied voltage, the 2D Cartesian streamers were up to four times thinner and slower, with the largest differences occurring near the start of the discharge.
Furthermore, their maximal electric field was about 30\% lower and their degree of ionization was about 65\% lower.
These differences in streamer properties can to some extent be explained by differences in the respective inception voltages.
For several electrode lengths and widths, we found inception voltages to be about twice as high in a 2D Cartesian geometry, due to the weaker electric field enhancement.

We therefore also performed a comparison at a similar ratio of applied voltage over inception voltage.
Velocities then became rather similar in the two types of models, and so did the streamer radii at later propagation times.
However, the maximal electric field in the 2D Cartesian case was still about 20-30\% lower, and the degree of ionization was about 40-50\% lower.

We have briefly looked at several relations between streamer properties, such as velocity, radius, maximal electric field and degree of ionization.
Furthermore, we have show that streamer branching cannot qualitatively be reproduced in a 2D Cartesian simulations.
Branching only occurs when strong noise is added to such simulations, and the resulting branches are much thinner than in real discharges in air.

Our findings can help to interpret the results of 2D Cartesian simulations, which can be a valuable tool to qualitatively study streamer phenomena under conditions that are computationally too expensive to simulate in full 3D.

\section*{References}.
\bibliographystyle{unsrt}
\bibliography{comparison-ref}

\end{document}